\documentclass[epj]{svjour}
\usepackage{latexsym}
\usepackage{graphics}
\begin{document}
\title{Synthesis, structural and physical properties of $\delta'$-FeSe$_{1-x}$}
\author{Mariano de Souza, Amir-Abbas Haghighirad, Ulrich Tutsch,
Wolf Assmus, and Michael Lang}

%
\offprints{mariano@physik.uni-frankfurt.de, a.a.haghi@physik.uni-frankfurt.de}          

\institute{Physikalisches Institut, Goethe-Universit\"at
Frankfurt(M), D-60438 Frankfurt am Main, Germany}

\date{Received: date / Revised version: date}

%
%
\abstract{
We report on synthesis, structural characterization, resistivity, magnetic and thermal expansion measurements  on the as yet unexplored $\delta'$-phase of FeSe$_{1-x}$, here synthesized under ambient- (AP) and high-pressure (HP) conditions. We show that in contrast to  $\beta$-FeSe$_{1-x}$, monophasic superconducting  $\delta'$-FeSe$_{1-x}$ can be obtained in off-stoichiometric samples with excess Fe atoms preferentially residing in the van der Waals gap between the FeSe layers. The AP $\delta'$-FeSe$_{1-x}$ sample studied here ($T_c$ $\simeq$ 8.5\,K) possesses an unprecedented residual resistivity ratio RRR  $\simeq$ 16. Thermal expansion data reveal a small feature around $\sim$90\,K, which resembles the anomaly observed at the structural and magnetic transitions for other Fe-based superconductors, suggesting that some kind of "magnetic state" is formed also in FeSe.
For HP samples (RRR $\simeq$ 3), the disorder within the FeSe layers is enhanced through the introduction of vacancies, the saturated magnetic moment of Fe is reduced and only spurious superconductivity is observed.
\PACS{\newline
      {74.62.Bf} {Effects of material synthesis, crystal structure, and chemical composition} \newline
      {72.15.Eb} {Electrical and thermal conduction in crystalline metals and alloys} \newline
      {74.25.Ha} {Magnetic Properties} \newline
      {65.40.De} {Thermal expansion}
     } 
} 
\maketitle
\section{Introduction}
The discovery of superconductivity in Fe-pnictide \cite{Kamihara08} with critical temperatures  $T_c$s as high as 55\,K attracted enormous interest to this class of materials.
Common to this new family of high-$T_c$ superconductors is a tetragonal structure at room temperature consisting of layers of Fe\emph{Pn} or Fe\emph{Ch} (\emph{Pn} and \emph{Ch} being a pnictogen or chalcogen atom, respectively), where Fe is tetrahedrally coordinated by \emph{Pn} or \emph{Ch} atoms. Prominent examples are LaFeAsO$_{1-x}$F$_x$ (``1111'') \cite{Kamihara08,chen08}, Ba$_{1-x}$K$_{x}$Fe$_2$As$_2$ (``122'') \cite{rotter08}, Li$_{1-x}$FeAs \linebreak  (``111'') \cite{Wang08}, and FeSe$_{1-x}$ (``11'') \cite{Hsu08}, see also \cite{Ishida,Johnston} for recent reviews. The $\beta$ polymorph of the FeSe phase exhibits superconductivity at ambient pressure with $T_c$ $\simeq$ 8.5\,K.
This compound possesses a layered anti-PbO tetragonal structure (space group \emph{P}4/\emph{nmm}) at room temperature. Upon cooling, a tetragonal-to-orthorhombic (space group \emph{Cmma}) transition takes place around 90\,K \cite{Mcqueen09}. Interestingly, in this compound, the structural transition is not followed or accompanied by a long-range magnetic ordering \cite{Imai09}, as  occurs in other Fe-based superconductors, see e.g.\,\cite{Jesche10} and references therein, although the presence of strong spin fluctuations for $T$ $\leq$ 100\,K has been pointed out in the literature \cite{Imai09}.
Due to the simplicity of its structure, i.e.\,the absence of guest ions or interleaved slabs in the van der Waals gap between the FeSe layers, FeSe$_{1-x}$ has been considered as an interesting target material for studying the intrinsic properties of Fe-based superconductors \cite{Buechner09}.
However, a relevant open issue concerns the composition of the superconducting phase.
For instance, in the original work \cite{Hsu08} reporting the discovery of superconductivity in FeSe$_{1-x}$, the authors claimed that the composition of the superconducting phase lies between FeSe$_{0.82}$ and FeSe$_{0.88}$ when these materials are synthesized at temperatures $\leq$ 700\,$^\circ$C. More recent studies, however, including various experimental techniques, such as neutron scattering as well as Electron Probe Microanalysis (EPMA), have shown that superconductivity in single-phase samples of FeSe$_{1-x}$ can only occur in the nearly stoichiometric compositions, i.e.\,close to 1:1 ratio of Fe to Se \cite{Mcqueen08,McqueenSSC09}. Several chemical substitution routes have been reported aiming to reach higher $T_c$ values in FeSe$_{1-x}$. On the one hand, attempts to increase $T_c$ by substituting  Fe by Ni, Co \cite{Mizuguchi09} and Cu \cite{Williams09} have proved unsuccessful, since superconductivity is destroyed in this process. On the other hand, substitution of  Se by S or Te results in a slight increase of $T_c$ \cite{Mizuguchi09}. Remarkably, application of a pressure of 8.9\,GPa increases $T_c$ from 8.5 to 36\,K \cite{Medvedev09}. Hence,  an obvious question that arises is whether the ``36\,K high-$T_c$'' crystal structure  could be stabilized by means of HP/high-temperature (HT)-synthesis. To the best of our knowledge,  there is currently only one literature report on FeSe$_{1-x}$ synthesized under HP conditions \cite{Sidorov}.

 There is no question that a high-temperature $\delta'$-FeSe phase of unknown structure exists \cite{Schuster79}, which melts congruently at 1075\,$^\circ$C. According to the phase diagram \cite{Schuster79,Svendsen72}, the $\delta'$-phase of FeSe$_{1-x}$ has a wide range of homogeneity between 48.5--62 at\% Se, while the $\beta$-phase is restricted to a narrow homogeneity range of 49--49.4\,at\% Se which appears below $\sim$450\,$^\circ$C \cite{Schuster79}.  Here we report on the synthesis, structural characterization, resistivity, magnetic and thermal expansion measurements on $\delta'$-FeSe$_{1-x}$ prepared under AP and HP conditions. Superconductivity was only found in the $\delta'$-FeSe$_{1-x}$ samples synthesized at AP conditions.  We focus basically on four points; we show that: i) similar to the other phases of FeSe, $\delta'$-FeSe$_{1-x}$ possesses a tetragonal structure at room temperature; ii) superconductivity in $\delta'$-FeSe$_{1-x}$ can be observed in off-stoichiometric (monophasic) samples; iii) for samples synthesized under HP conditions, disorder within the FeSe layers is enhanced and only spurious superconductivity is observed; iv) the thermal expansion coefficient for AP $\delta'$-FeSe$_{1-x}$ shows an anomaly around 90\,K similar to that found in the "1111" and "122" families at their structural and magnetic transitions.

\section{Experimental Details}
The samples were prepared from 99.99$+$\% Fe (Johnson, Matthey  Co., Ltd; turning from rods) and 99.999\% Se (Chempur, granules of 2--4\,mm). To prepare the alloys, equiatomic quantities of selenium and iron lumps were weighed into silica ampoules which were evacuated and sealed. A total amount of starting material of approximately 3\,g was melted by induction and chill cast using a high-frequency (HF) generator (300\,kHz). This method relies on rapid melting and chilling to minimize the loss of Se as vapor and pronounced alteration in the composition of the alloy \cite{Nesmeyanov}.
Rapid cooling ($\sim$\,20$^\circ$C/s at HT) of the sample from the melt was done by pulling out the ampoule of the the HF-generator coil and dropping it on sand.
The samples were kept in vacuum desiccators before the measurements were taken.
Powders of $\delta'$-FeSe$_{1-x}$ synthesized at AP conditions were used as the precursor material for the HP/HT synthesis in a 6--8 type multianvil apparatus \cite{Haghi}. The precursor material was put in a BN crucible and heated up to $\sim$1200\,$^\circ$C under a pressure of $\sim$\,5.5\,GPa and kept at these conditions for 30\,min before quenching to room temperature.
The crystal structure of the samples was refined at room temperature by X-ray powder diffraction (XRD) measurements using Cu-K$_{\alpha}$ radiation. The analysis of the diffraction data was performed with the GSAS suite of Rietveld programs \cite{Larsen}. A micro-structural analysis was carried out by using EPMA.
This analysis, together with magnetic susceptibility measurements on various samples taken from the original batch,
revealed the presence of superconducting and non-superconducting specimens (see below) as well as the presence of non-superconducting wire-like fibers with a diameter of $\sim$100$\,\mu$m, see also \cite{Pimentel}.
For the present study, five samples were used:\newline
\begin{itemize}
\item Sample $\#$1,  $\delta'$-FeSe$_{0.95}$, AP, starting composition Fe/Se 1:1
\item Sample $\#$2,  $\delta'$-FeSe$_{0.96}$, AP, starting composition Fe/Se 1:1
\item Sample $\#$3,  $\delta'$-Fe$_{0.95}$Se, HP (sample $\#$1 was used as precursor for HP synthesis)
\item Sample $\#$4,   $\delta'$-FeSe$_{0.96}$ containing non-magnetic $\gamma$-Fe \cite{Tisza,Note} dendrites as inclusions, AP, starting composition Fe/Se 1:0.88
\item Sample $\#5$,   $\delta'$-Fe$_{0.91}$Se containing  $\gamma$-Fe \cite{Tisza}, HP (synthesized by using the elements Fe and Se as starting materials), starting composition Fe/Se 1:1.
\end{itemize}
Susceptibility and magnetization measurements were performed using a Quantum Design SQUID magnetometer (Quantum Design
Magnetic Property Measurement System).
The electrical resistance was measured by employing a standard four terminal
ac technique operating at 16\,Hz. Measurements on cooling and warming  have been taken by employing a sweep rate of $\pm$6\,K/h. A $^4$He-gas-pressure technique was
used to study the resistance under hydrostatic-pressure conditions. The  pressure changes were performed at room temperature and recorded using an InSb resistance as pressure gauge.
The thermal expansion coefficient,
$\alpha(\textit{T})=\textit{l}^{-1}(\partial \textit{l}/\partial
\textit{T})$, was measured by employing an ultra-high-resolution
capacitance dilatometer \cite{Pott} with a maximum resolution of $\Delta
l/l=10^{-10}$.

\section{Results and Discussions}

\subsection{Structural and Chemical Analysis}
\begin{figure}
\centering
\resizebox{0.37\textwidth}{!}{%
  \includegraphics{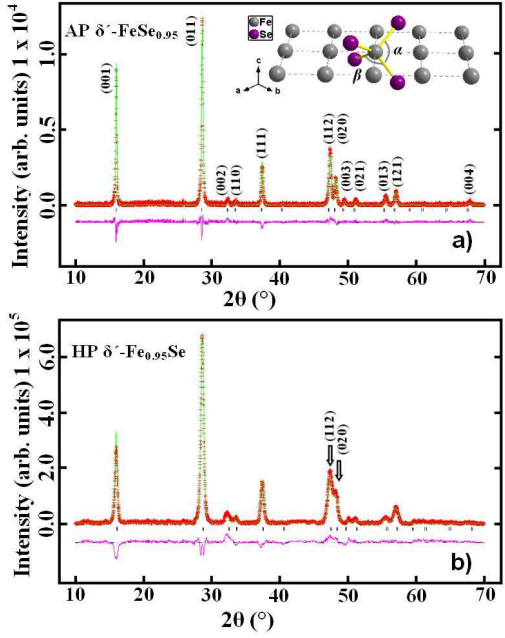}}
\caption{(Color online) Room temperature X-ray powder diffraction data (2$\theta$ = 10$^\circ$ -- 70$^\circ$) for a) AP $\delta'$-FeSe$_{0.95}$ ($\#$1) and b) HP $\delta'$-Fe$_{0.95}$Se ($\#$3). Miller indices are indicated on the apex of the reflections. The lower base-lines indicate the difference between the calculated and the measured X-ray intensities. Inset:  Cut out of the tetragonal FeSe$_{1-x}$ crystal structure consisting of layers of FeSe$_4$ edge-sharing tetrahedra. }\label{DP}
\end{figure}

Fig.\,\ref{DP} shows the XRD patterns for samples $\#1$ and $\#3$ synthesized at AP and HP conditions, respectively. Two phases have been observed at room temperature: (1)  $\delta'$-Fe$_{1.05}$Se (= $\delta'$-FeSe$_{0.95}$) designated in Fig.\,\ref{DP}a as AP $\delta'$-FeSe$_{0.95}$. The latter is the phase with excess of iron atoms in interstitial positions; (2) The HP/HT $\delta'$-Fe$_{0.95}$Se phase with iron vacancies designated in Fig.\,\ref{DP}b as HP $\delta'$-Fe$_{0.95}$Se. Both phases possess an anti-PbO tetragonal structure \linebreak (space group \emph{P}4/\emph{nmm}) at room temperature, reported here for the first time, although the existence of the $\delta'$ phase was already reported in Ref.\,\cite{Schuster79}.
The summary of the structure refinements is listed in Table\,1. These include the unit cell parameters, atomic positions, and isotropic displacements. The refinements were  performed using the following choice of the crystallographic Wyckoff-positions with Fe in (2\emph{b}) position ($\frac{1}{4}$, $\frac{3}{4}$, $\frac{1}{2}$) and Se in (2\emph{c}) position ($\frac{1}{4}$, $\frac{1}{4}$, \emph{z}). The two samples,  AP $\delta'$-FeSe$_{0.95}$ ($\#$1) and HP $\delta'$-Fe$_{0.95}$Se ($\#$3), were chemically analyzed by EPMA.

\begin{figure}
\centering
\resizebox{0.33\textwidth}{!}{%

 \includegraphics{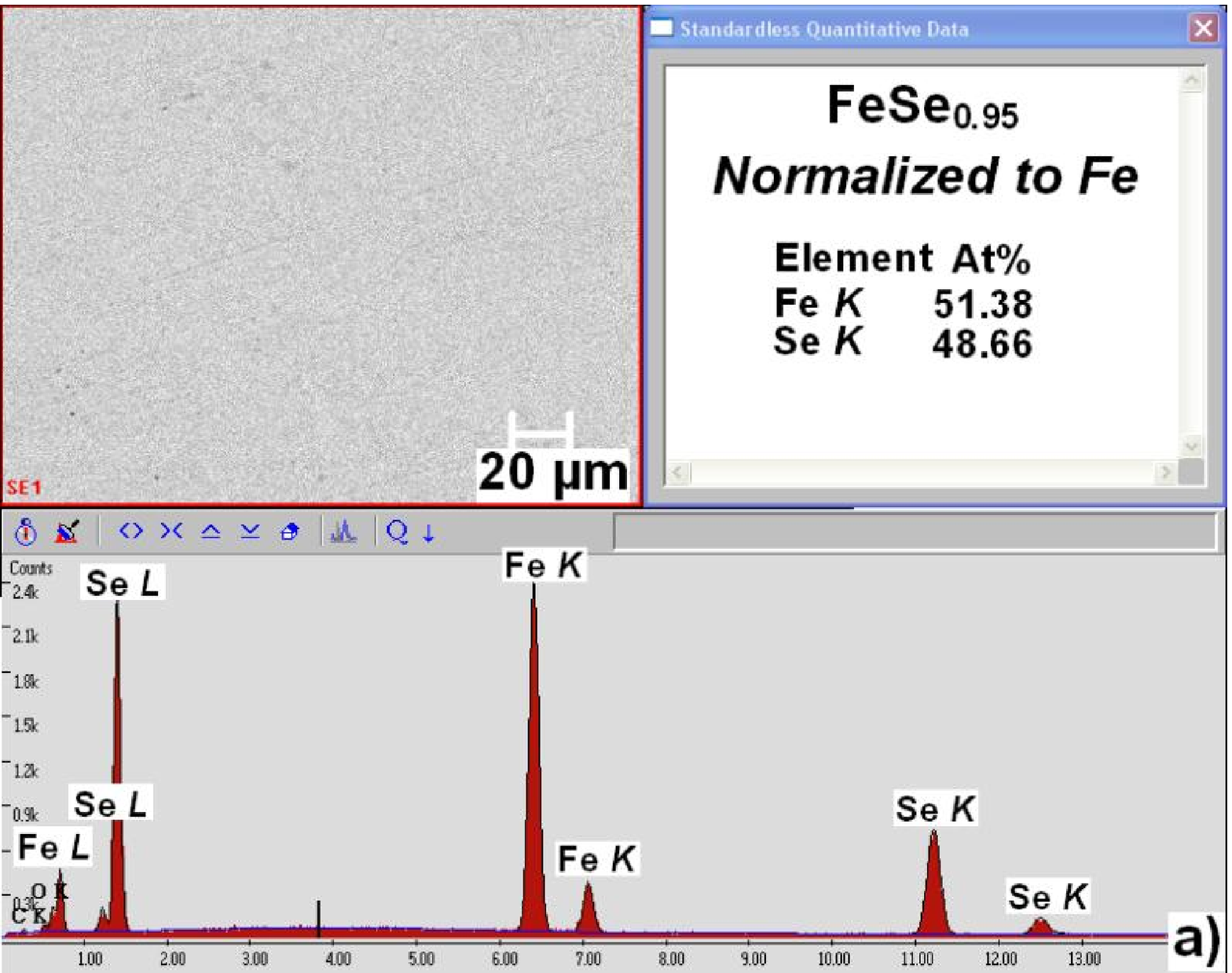}}
\vspace{0.5cm}

\resizebox{0.33\textwidth}{!}{%
 \includegraphics{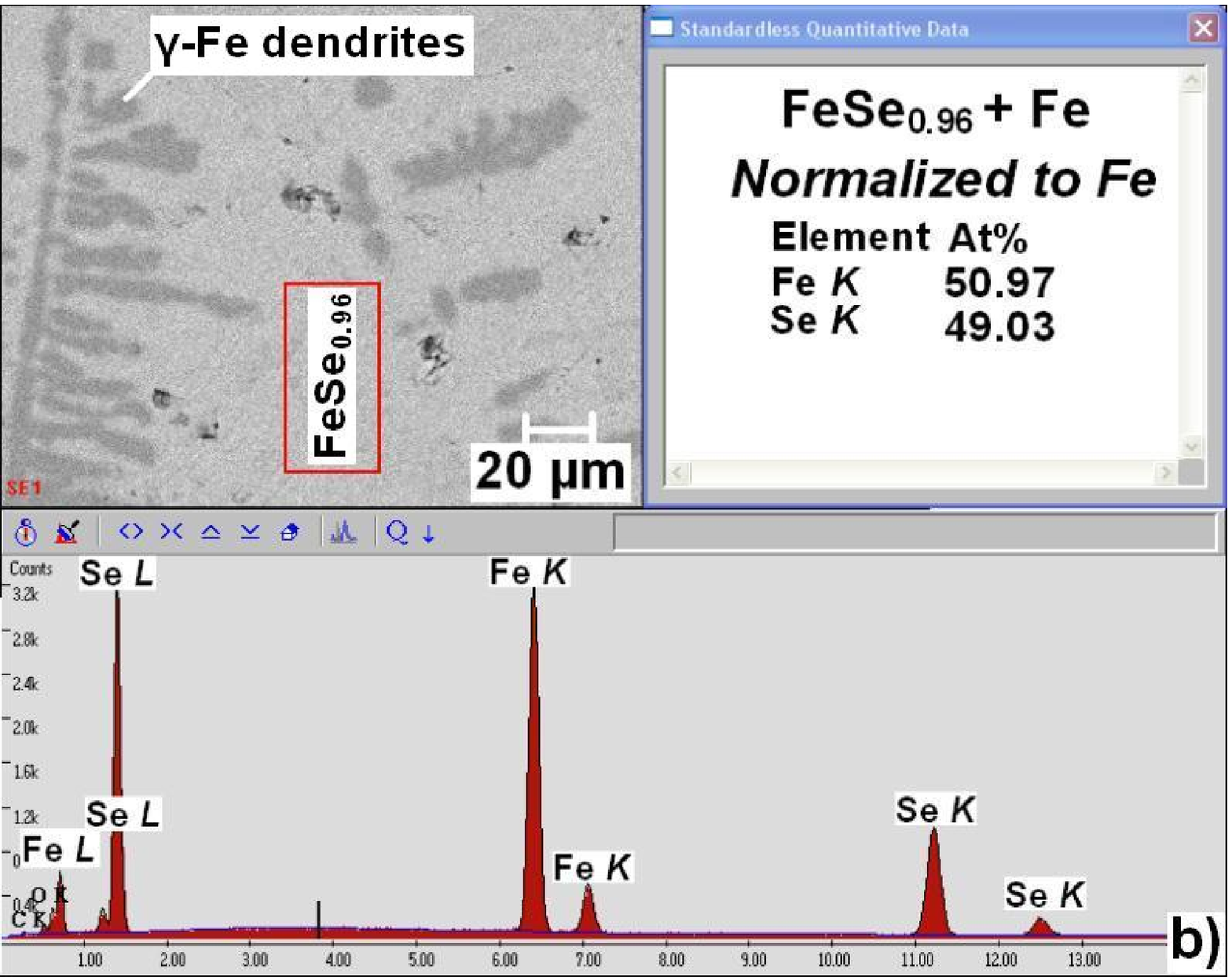}}
\caption{(Color online) Scanning electron microscopy  micrographs and EPMA patterns of $\delta'$-FeSe$_{1-x}$ prepared at ambient-pressure conditions, (a) monophasic AP $\delta'$-FeSe$_{0.95}$ ($\#$1), and (b) AP $\delta'$-FeSe$_{0.96}$ containing $\gamma$-Fe dendrites ($\#$4).}\label{EDX}        
\end{figure}
On the iron-rich side, EPMA and X-ray (not shown here) analyses  of samples quenched from $\sim$700\,$^\circ$C revealed two phases,  namely $\beta$-FeSe and $\alpha$-Fe, also reported by McQueen \emph{et al.} \cite{McqueenSSC09}.
The diffraction pattern of HP $\delta'$-Fe$_{0.95}$Se  highlights a representative peak shift of all reflections compared to the AP $\delta'$-FeSe$_{0.95}$ diffraction data (Fig.\ref{DP}a), which is consistent with a lattice contraction accompanied by a progressive incorporation of vacancies in the FeSe layers upon pressurizing the lattice.
As can be seen from Fig.\,\ref{DP}, for both samples the (011) reflexion peak predominates. On comparing their diffraction patterns, one can see that all reflections are broadened and the intensities of some reflections, e.g.\,(001), (013), and (004),  are reduced for the HP $\delta'$-Fe$_{0.95}$Se sample. Furthermore, while the (112) and (020) reflections in the AP $\delta'$-FeSe$_{0.95}$ diffraction pattern are split, both peaks move close to each other in HP $\delta'$-Fe$_{0.95}$Se. The latter is a direct consequence of the lattice contraction  of 2.1\% (see Table\,\ref{Table}).

\begin{table}
\centering
\caption{\small  Refined structural parameters of  AP
$\delta'$-FeSe$_{0.95}$ and HP $\delta'$-Fe$_{0.95}$Se with
selected bond lengths and angles obtained from Rietveld refinements
of XRD-data collected at room temperature. The Bragg \emph{R} factor
is given for the main phase; the estimated errors in the last digits
are given in  parentheses. $^\ast$For the angles $\alpha$ and
$\beta$, see inset of Fig.\,\ref{DP}a.}\label{Table}
\small
\begin{tabular}{p{2.5cm} p{2.5cm} p{2.5cm}}
\hline\hline\noalign{\smallskip}
\textbf{Sample} & \textbf{AP $\delta'$-FeSe$_{0.95}$ ($\#1$)} & \textbf{HP $\delta'$-Fe$_{0.95}$Se ($\#3$)}\\
\noalign{\smallskip}\hline\noalign{\smallskip}
 $T_c$(K) & \small 8.5 & \small Spurious SC\\
 Space group & \small \emph{P4/nmm} & \small \emph{P4/nmm}\\
 Refined composition & \small FeSe$_{0.95}$ (Fe$_{1.05}$Se) & \small Fe$_{0.95}$Se  \\
 \emph{a} (${\AA}$) & \small 3.7891(2) & \small 3.7647(2) \\
 \emph{c} (${\AA}$) & \small 5.5478(5) & \small 5.5048(9)\\
 \emph{V} (${\AA^3}$) $-$ \emph{c}/\emph{a} & \small 79.655 $-$ 1.4641 & 78.023 $-$ 1.4622\\
 Se \emph{z} & \small 0.2322 & \small 0.2320 \\
 Occupancy & \small Se-0.9533, Fe-1 & \small Fe-0.9450, Se-1 \\
 B$_{iso}$($\AA^2$) & \small 0.11(3) & \small 0.084(5)  \\
 $R_{wp}$ & \small 3.45 & \small 5.42 \\
 $R_{exp}$ & \small 2.49 & \small 4.23 \\
 $\chi^2$ & \small 1.92 & \small 1.64 \\
 Fe-Se ($\AA$) & \small 2.4076 & \small 2.3916 \\
 Fe-Se-Se ($^\circ$) & \small $\alpha$ $-$ 103.793$^\ast$ &  \small $\alpha$ $-$ 103.825$^\ast$ \\
    & \small $\beta$ $-$ 112.383$^\ast$ &  \small $\beta$ $-$ 112.366$^\ast$ \\
 Se-Se ($\AA$) & \small 3.7891 & \small 3.7647 \\
 Se-Se ($\AA$) & \small 4.0010 & \small 3.9740 \\
 mag.\,moment ($\mu_B$/Fe)   & \small 0.33 & \small 0.21\\
 at 5\,T/150\,K  &  & \\ \hline\hline
\end{tabular}
\end{table}

Our chemical analyses revealed that the AP synthesis described above (see discussion below for $\#1$ and $\#2$), produces single-phase samples of the main composition FeSe$_{0.95 \pm 0.02}$. As can be seen from Fig.\,2a, EPMA mapping data for Fe and Se confirm the uniform distribution of both elements throughout the sample ($\#1$). However, the  increase of the Fe content in the starting composition (Fe:Se = 1:0.88) in the AP synthesis revealed the presence of AP $\delta'$-FeSe$_{0.96}$ and an excess amount of $\gamma$-Fe ($\#$4)  that formed a dendritic structure as a consequence of rapid crystallization (see Fig.\,2b). We stress that the chemical analysis of the HP-$\delta'$-Fe$_{0.95}$Se ($\#$3) revealed 10\,\% Fe deficiency compared with the starting composition. This is consistent with the observation of a thin layer of Fe on the BN crucible wall after the synthesis.
The compositions found by EPMA were used in the structural refinements of the X-ray diffraction data (e.g.\,the Se occupancy and the atomic displacement parameters).
The results of the chemical and the structural analyses  discussed above are in perfect agreement with studies of the FeSe phase diagram reported in the literature  \cite{Schuster79,Svendsen72}. Note that for samples $\#$1, $\#$2 and $\#$3, no indication for foreign phases  could be detected. Low-temperature structural data analysis is underway and will be published elsewhere.

\subsection{Physical Properties}
\begin{figure}
\centering
\resizebox{0.47\textwidth}{!}{%
  \includegraphics{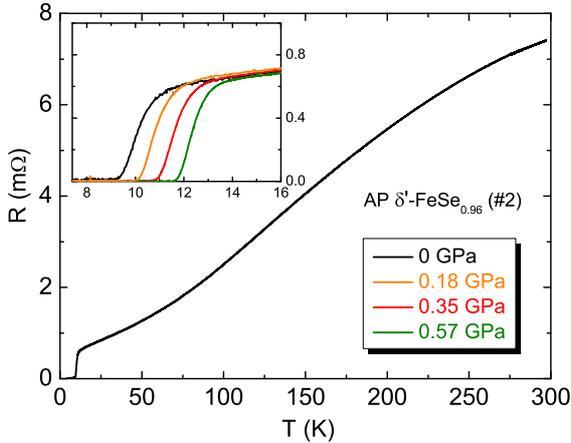}
}
\caption{(Color online) Main panel: Resistance data for AP
$\delta'$-FeSe$_{0.96}$ ($\#$2) under ambient pressure. Inset:
blowup of the low-temperature data under selected pressures, as
indicated in the label.}
\label{R}       
\end{figure}

\begin{figure}
\centering
\resizebox{0.50\textwidth}{!}{%
  \includegraphics{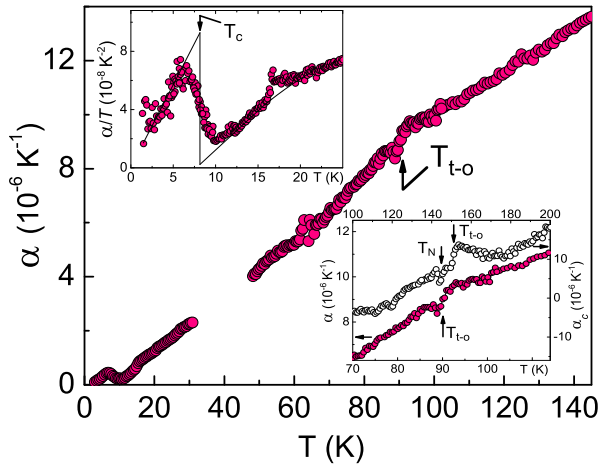}}
\caption{(Color online) Main panel: Thermal expansion coefficient
$\alpha$($T$) for  AP $\delta'$-FeSe$_{0.95}$ ($\#$1). $T_{t-o}$
indicates the tetragonal-to-orthorhombic transition temperature,
determined in Ref.\,\cite{Mcqueen09}. Missing data in the $T$ window
35 -- 45\,K is due to an enhanced noise level in this $T$ range.
Inset (upper left): Blowup of the data below 25\,K in an
$\alpha$/$T$ versus $T$ plot. Straight lines represent an
``equal-areas'' construction to determine $T_c$ (8.5\,K). Inset
(lower right):  $\alpha$ versus $T$ on an expanded temperature scale
close to the
structural phase transition (bottom/left scale). For comparison, $\alpha_c$ versus $T$ for single-crystalline CeFeAsO is shown (top/right scale) with $T_N$ indicating the magnetic ordering of Fe (taken from Ref.\,\cite{Jesche10}). Solid line is a guide for the eyes.}\label{te}       
\end{figure}

In Fig.\,\ref{R} we show resistance data for single-phase AP
$\delta'$-FeSe$_{0.96}$ ($\#$2) under ambient and selected
$^4$He-gas pressures  in the temperature range 2--300\,K. The
measurements were conducted in five consecutive runs in the
following sequence: ambient pressure, 0.35\,GPa,  0.57\,GPa  (after
this run pressure was removed completely) and a final run under
0.18\,GPa. Within the pressure range studied here, AP
$\delta'$-FeSe$_{0.96}$  presents a metallic behavior
(d$R$/d$T$ $>$ 0) down to low temperatures with a residual
resistivity ratio RRR = $R$(300\,K)/$R_0$ = 16. $R_0$ was determined
by extrapolating the normal-state resistance to $T$ = 0 using a
second-order polynomial function. To our knowledge, a value RRR = 10
for $\beta$-Fe$_{1.01}$Se is the highest residual resistivity ratio
to the present date reported in the literature \cite{Mcqueen08}.
Thus, the observed RRR of 16 reflects the
low defect concentration of the sample studied here. Under
ambient-pressure conditions,
the data reveal the superconducting transition (zero resistance)
around 9.0\,K. The observation of superconductivity in AP
$\delta'$-FeSe$_{0.96}$ brings into question the proposal
\cite{McqueenSSC09} that only samples of FeSe$_{1-x}$ close to the
1:1 stoichiometry show superconductivity (see below).  Upon applying
hydrostatic pressure, $T_c$ shifts to higher temperatures (see inset
of Fig.\,\ref{R}), as reported by other groups
\cite{Medvedev09,Masaki09,Margadonna09}.
Using  ``$R$ = 0'' as the criterium for the superconducting
transition temperature, we obtain $T_c\mid_{0.18\,GPa}$ = 9.9\,K,
$T_c\mid_{0.35\,GPa}$ = 10.7\,K, and $T_c\mid_{0.57\,GPa}$ =
11.6\,K. Assuming a linear increase of $T_c$ as pressure is
increased, one obtains d$T_c$/d$P_{hydr}$ = $+$(4.3$\pm$0.5)\,K/GPa
for $P_{hydr}$ $\leq$ 0.57\,GPa, the highest pressure applied in our
experiments. This pressure coefficient \cite{pdep} is somehow higher
than that obtained using a diamond anvil cell (d$T_c$/d$P$ =
$+$1.4\,K/GPa) \cite{Garbarino09}, but lower than the one obtained
using a clamped piston-cylinder cell with Fluorinert as a
pressure-transmitting medium (d$T_c$/d$P$ = $+$9.1\,K/GPa)
\cite{Mizuguchi}. This discrepancy might be associated with
differences in the sample quality, and the variation of
composition, as pointed out in Ref.\,\cite{Huang09}, as well as
with the non-hydrostatic conditions inherent to pressure experiments
employing oil as a pressure-transmitting medium. These results imply
that hydrostatic conditions are not a prerequisite for the increase
of $T_c$ under pressure in FeSe$_{1-x}$, in contrast to the
situation encountered in CaFe$_2$As$_2$, where it appears that superconductivity can
be induced only under  non-hydrostatic conditions \cite{Yu09}.
\begin{figure}
\centering
\resizebox{0.49\textwidth}{!}{%
  \includegraphics{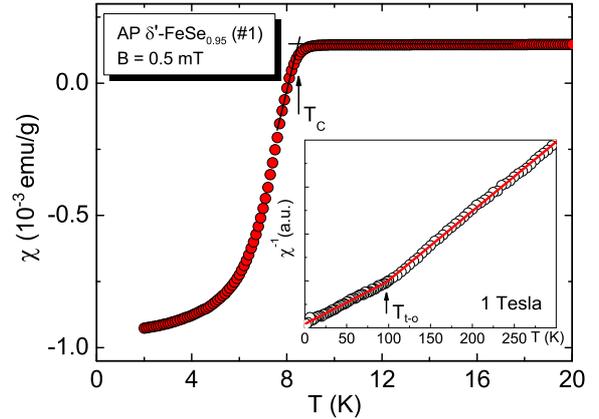}}
\caption{(Color online) Main panel: Magnetic susceptibility for AP
$\delta'$-FeSe$_{0.95}$ ($\#$1). Data were taken at a small magnetic
field of 0.5\,mT. The superconducting transition temperature $T_{c}$ corresponds to the temperature at which the linear extrapolations of $\chi(T)$ from above and below cross. Inset: Inverse  magnetic susceptibility for sample
$\#$1 at 1\,T. Solid straight lines are linear fits, indicating two distinct
magnetic regimes, cf.\,discussion in the main text. $T_{t-o}$
indicates the tetragonal-to-orthorhombic transition temperature,
determined in Ref.\,\cite{Mcqueen09} for $\beta$-FeSe$_{1-x}$.}
\label{chi}      
\end{figure}
In the main panel of Fig.\,\ref{chi}, we present the
temperature-dependent magnetic susceptibility below 20\,K for AP
$\delta'$-FeSe$_{0.95}$ ($\#$1). The data, taken upon cooling in
a magnetic field of 0.5\,mT, reveal clearly the
superconducting transition at $T_c$ = 8.5\,K.
Thermal expansion experiments on AP $\delta'$-FeSe$_{0.95}$ ($\#$1)
are shown in Fig.\,\ref{te}. It is worth mentioning here that
thermal expansion measurements on polycrystalline samples quantify
the directional average value for the uniaxial thermal expansion coefficients
$\alpha_i$($T$), with \emph{i} = \emph{a}, \emph{b} and
\emph{c}-axes,  which precludes any directional-dependent analysis.
In other words, thermal expansion on polycrystalline samples probes
the volumetric expansivity $\beta$, i.e.\,$\alpha$ $\simeq$
$\beta$/3.
As can be seen from Fig.\,\ref{te}, upon cooling, $\alpha$($T$)
decreases almost linearly with a change of slope around  90\,K.
The small but distinct anomaly visible here is attributed to the tetragonal-to-orthorhombic
structural transition, revealed at $T_{t-o}$ = 90\,K also for $\beta$-FeSe$_{1-x}$ \cite{Mcqueen09}.
The shape of the $\alpha$($T$) anomaly (cf.\, lower right inset of Fig.\,\ref{te}), i.e.\,a small maximum at the high-$T$ side followed by a rapid reduction and a dip, resembles the anomalies observed at $T_{t-o}$ in related materials. This includes polycrystalline samples of undoped LnFeAsO (Ln = La, Ce, Sm, Gd) \cite{Klingeler09} as well as single crystals of Co-doped BaFe$_2$As$_2$
\cite{Budko09} and CeFeAsO \cite{Jesche10} measured along the \emph{c}-axis. The latter data are shown, for comparison, in the lower right inset of Fig.\,\ref{te}.
The striking similarity of the $\alpha$($T$)-anomaly around $T_{t-o}$ for
$\delta'$-FeSe$_{1-x}$ to those observed in the undoped or underdoped variants of the various other Fe-based superconductors is remarkable, as in all these systems -- except FeSe$_{1-x}$ -- the structural transition is
accompanied or followed by the onset of long-range magnetic order at $T_N$
with $T_N$ $\leq$ $T_{t-o}$, c.f. the closely spaced transitions at $T_N$ and $T_{t-o}$ in CeFeAsO shown in the lower right inset of Fig.\,\ref{te} \cite{Jesche10}. This may suggest that also in FeSe some kind of "magnetic state" forms slightly below $T_{t-o}$.
In fact, a closer inspection of the magnetic susceptibility $\chi$, shown in the inset of Fig.\,\ref{chi} as $\chi^{-1}$ vs. $T$, reveals a change in the magnetic behavior around $T_{t-o}$. It has to be shown by sensitive microscopic probes, such as neutron scattering, whether the break in the slope of $\chi^{-1}(T)$ around $T_{t-o}$ actually signals a cooperative magnetic effect, or whether it simply reflects changes in the magnetic interactions as a consequence of the structural changes associated with $T_{t-o}$.
\begin{figure}
\centering
\resizebox{0.48\textwidth}{!}{%
  \includegraphics{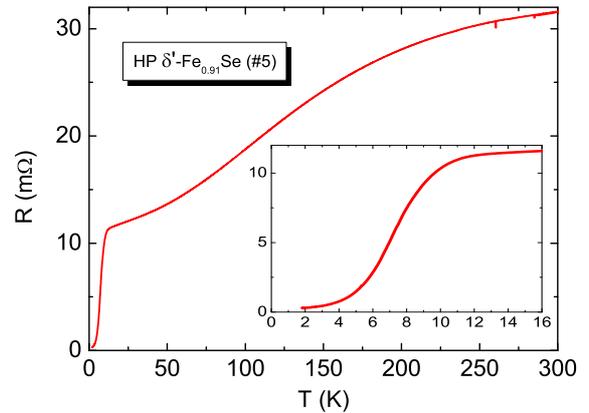}
}
\caption{(Color online) Main panel: Resistance data for HP
$\delta'$-Fe$_{0.91}$Se ($\#$5). Inset: Blowup of the
low-temperature data.}\label{RHP}      
\end{figure}
At lower temperatures, $\alpha$($T$) shows a shallow maximum around 17\,K (see
left upper inset of Fig.\,\ref{te}). Remarkably, this feature
coincides with the temperature at which X-ray diffraction experiments on $\beta$-FeSe$_{0.99}$
 revealed that the torsional
angle between the Se pairs changes from 90$^\circ$ to 89.7$^\circ$
\cite{Mcqueen09}. As a consequence, one
Fe-Fe distance in $\beta$-FeSe$_{0.99}$ shrinks, while the second
Fe-Fe distance elongates, resulting in an average change in the
short-long Fe-Fe separation of 0.012\,${\AA}$ \cite{Mcqueen09}.
Thus, we attribute the signature in $\alpha$($T$) around 17\,K to
the above-mentioned structural changes in the local surrounding of
Fe. The superconducting transition is clearly observed in our
thermal expansion experiments. The discontinuity in $\alpha$ around 8.5\,K with $\Delta \alpha \propto \Delta \beta >$ 0,
indicates (upper left inset of Fig.\,\ref{te}), according to the
Ehrenfest relation, d$T_c$/d$P$ $\propto$ $\Delta
\beta$/$\Delta C$, a positive hydrostatic pressure dependence of
$T_c$. The latter result is consistent with our resistivity data
under $^4$He-gas pressure, cf. Fig.\,\ref{R}.
In contrast to  AP $\delta'$-FeSe$_{1-x}$, samples synthesized under HP/HT-conditions revealed only traces of superconductivity. As can be seen in the inset of Fig.\,\ref{RHP}, the resistance of HP $\delta'$-Fe$_{0.91}$Se ($\#$5) remains finite
down to low temperatures and no diamagnetic signal was observed (not shown). 
\begin{figure}
\centering
\resizebox{0.48\textwidth}{!}{%
  \includegraphics{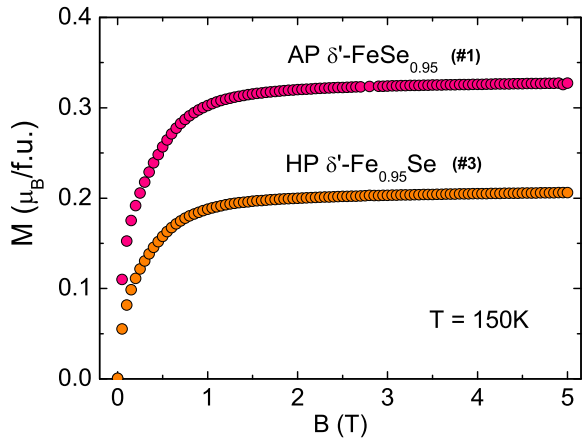}
}
\caption{(Color online) Magnetization curves at 150\,K for samples
AP $\delta'$-FeSe$_{0.95}$ ($\#$1) and HP $\delta'$-Fe$_{0.95}$Se
($\#$3) (see main text).}
\label{mag}     
\end{figure}
Although we cannot definitely rule out that spurious superconductivity is the result of traces of superconducting $\delta'$-FeSe with $T_c$ of $\sim$8.5\,K, the midpoint of the very broad transition around 8\,K, which is lower than that of superconducting $\delta'$-FeSe,  points against such a scenario. According to the pressure dependence of $T_c$, obtained in the inset of  Fig.\,\ref{R}, we would expect that the pressure exerted by the matrix of the HP phase (due to the reduced lattice parameters) on the $\delta'$-FeSe precipitations, should give rise to an increase of $T_c$.

An important question, which has not yet been answered satisfactorily, regards the magnetic background and its relation to the superconducting state in FeSe$_{1-x}$.
To this end, magnetization measurements were conducted at 150\,K
from 0 to 5\,T for samples AP  $\delta'$-FeSe$_{0.95}$ ($\#$1) and HP $\delta'$-Fe$_{0.95}$Se
($\#$3). As Fig.\,\ref{mag} indicates, we find a rapidly saturating magnetization with distinctly different values for the saturation magnetization $M_{s}$ corresponding to 0.33 and 0.21 $\mu_{B}$/(Fe atom) for the AP ($\#$1) and HP ($\#$3) samples, respectively. Similar magnetization curves have been observed in FeSe$_{1-x}$ also by other groups \cite{McqueenSSC09,Patel09}. In Ref.\,\cite{McqueenSSC09} this magnetization has been related to Fe precipitations, verified also in the diffraction data on the samples studied there \cite{McqueenSSC09}. As X-ray and EPMA analyses gave no indication for traces of foreign phases in the present materials, the large magnetic moments observed here have to be of different origin. It is still premature to definitely state that a large magnetic moment is a necessary condition for the material to become superconducting at $T_{c}$ = 8.5 K. Yet, the substantially reduced moment by $\sim$35\% for the non-superconducting HP sample ($\#$3) would point in the same direction.
The Fe deficiency revealed for $\delta'$-Fe$_{0.95}$Se
($\#$3) after the HP/HT treatment of $\delta'$-FeSe$_{0.95}$ ($\#$1) is likely to be caused by the introduction of vacancies in the FeSe layers. In addition, those interstitial Fe atoms which are weaker bonded compared to the regular Fe atoms in the lattice, are likely to be expelled from the bulk as a consequence of HP/HT annealing. This is consistent with the reduction of the RRR value from
16 ($\#$1) to $\sim$3 ($\#$3) (see main panel of Fig.\,\ref{RHP}), and the broadening of the diffraction peaks for HP $\delta'$-Fe$_{0.95}$Se ($\#$3) (Fig.\,\ref{DP}b), and indicates
that superconductivity in FeSe is very sensitive to disorder in the FeSe layers.

\section{Conclusion}
To summarize, we have synthesized the $\delta'$ phase of FeSe under
ambient- and high-pressure conditions. The room-temperature
structure of AP $\delta'$-FeSe was refined for the first time. Our
analysis reveals that superconductivity in monophasic AP
$\delta'$-FeSe$_{1-x}$ can be observed in off-\linebreak stoichiometric
samples with excess Fe atoms preferentially residing in the van der Waals gap between the FeSe layers. The question whether this is a prerequisite for superconductivity, as a consequence of the generation of charge carriers, is unclear at present. We showed that high-pressure synthesis results in Fe-deficient samples due to the replacement of the
interlayer/interstitial vacancies by Fe atoms. This is accompanied by disorder in the FeSe layers, a reduction of the average magnetic moment and the occurrence of only spurious superconductivity.
In addition, we showed that the
thermal expansivity in the immediate vicinity of the
tetragonal-to-orthorhombic structural phase transition in
$\delta'$-FeSe$_{1-x}$ possesses a striking similarity with
literature results on other Fe-based materials, where the
structural transition is followed by a magnetic ordering of the Fe
magnetic moments. Finally we emphasize that in order to exploit the potential of the high-pressure synthesis, the influence of the annealing conditions, i.e. temperature and pressure, on the stabilization of the $\delta'$ phase and its superconducting properties has to be investigated in more detail. We anticipate that by the application of higher pressures during the synthesis, one should be able to stabilize an orthorhombic crystal symmetry with lattice parameters comparable to those characterizing the superconducting phase \cite{Marga08}.

\section{Acknowledgements}
This work is part of the DFG priority program  (SPP 1458) ``High-Temperature Superconductivity in Iron Pnictides'' funded by the German  Science Foundation.


\begin{thebibliography}{Buravov 271}
\bibitem{pdep} Using the midpoint as the superconducting transition temperature, we obtained a pressure coefficient of d$T_c$/d$P_{hydr}$ = $+$(3.9$\pm$0.5)\,K/GPa.
\end{thebibliography}
\end{document}